\documentclass[apjl]{emulateapj}
\usepackage{apjfonts}  %For emulateapj

\newcommand{\be}{\begin{equation}}
\newcommand{\ee}{\end{equation}}
\newcommand{\ba}{\begin{eqnarray}}
\newcommand{\ea}{\end{eqnarray}}
\newcommand{\Mc}{{\cal M}}
\newcommand{\Ms}{M_{\odot}}
\def\ltsima{$\; \buildrel < \over \sim \;$}
\def\simlt{\lower.5ex\hbox{\ltsima}}
\def\gtsima{$\; \buildrel > \over \sim \;$}
\def\simgt{\lower.5ex\hbox{\gtsima}}
\shorttitle{Spinning Inspiral Binaries and Gravitational-Wave Astronomy}
\shortauthors{M.V.\ van der Sluys et al.}

\begin{document}

\title{Gravitational-Wave Astronomy with Inspiral Signals of Spinning Compact-Object Binaries}

\author{M.V.\ van der Sluys$^{1}$, C.\ R\"{o}ver$^{2,3}$, A.\ Stroeer$^{1,4}$, V. Raymond$^{1}$, I.\ Mandel$^{1}$, \\
  N.\ Christensen$^{5}$, V.\ Kalogera$^{1}$, R.\ Meyer$^{2}$, A.\ Vecchio$^{1,4}$}
\altaffiltext{1}{Physics \& Astronomy, Northwestern U., Evanston IL, USA} 
\altaffiltext{2}{Statistics, U.\ of Auckland, Auckland, New Zealand}
\altaffiltext{3}{Max-Planck-Institut f\"{u}r Gravitationsphysik, Hannover, Germany}
\altaffiltext{4}{Physics \& Astronomy, U. of Birmingham, Edgbaston, Birmingham, UK}
\altaffiltext{5}{Physics \& Astronomy, Carleton College, Northfield  MN, USA}

\begin{abstract}
  Inspiral signals from binary compact objects (black holes and neutron stars) are primary 
  targets of the ongoing searches by ground-based gravitational-wave interferometers 
  (LIGO, Virgo, GEO-600 and TAMA-300). 
  We present parameter-estimation simulations for inspirals of black-hole--neutron-star 
  binaries using Markov-chain Monte-Carlo methods. 
  For the first time, we have both estimated the parameters of a binary inspiral source 
  with a spinning component and determined the accuracy of the parameter estimation,
  for simulated observations with ground-based gravitational-wave detectors.
  We demonstrate that we can obtain the distance, sky position, and binary orientation
  at a higher accuracy than previously suggested in the literature.
  For an observation of an inspiral with sufficient spin and two or three detectors
  we find an accuracy in the determination of the sky position of typically a few tens
  of square degrees.
\end{abstract}


\section{Introduction}
\label{sec:intro}

Binary systems with compact objects --- neutron stars (NS) and black holes (BH) --- in the mass 
range $\sim 1\,\Ms - 100\,\Ms$ are among the most likely sources of gravitational waves (GWs) 
for ground-based laser interferometers currently in operation (Cutler \& Thorne, 2002): LIGO 
(Barish \& Weiss 1999), Virgo (Arcese et al.\ 2004), GEO-600 (Willke et al.\ 2004) and TAMA-300 
(Takahashi et al.\ 2004). 
Merger-rate estimates are quite uncertain and for BH-NS binaries current detection-rate 
estimates reach as high as 0.1\,yr$^{-1}$ (\emph{e.g.} O'Shaughnessy et al.\ 2008) for first-generation 
instruments. 
Upgrades to Enhanced LIGO/Virgo (2008--2009) and Advanced LIGO/Virgo (2011--2014)
are expected to increase detection rates by factors of about $\sim 8$ and $10^3$, respectively. 

The measurement of astrophysical source properties holds major promise for 
improving our physical understanding and requires reliable methods for parameter estimation. 
This is a challenging problem because of the large 
number of parameters ($> 10$) and the presence of strong correlations among them, leading 
to a highly-structured parameter space. In the case of high mass ratio binaries (\emph{e.g.} BH-NS),
these issues are amplified 
for significant spin magnitudes and large spin misalignments 
(Apostolatos et al.\ 1994; Grandcl{\'e}ment et al.\ 2003; Buonanno et al.\ 2003). 
However, the presence of  
spins benefits parameter estimation through the signal modulations, although still  
presenting us with a considerable computational challenge. This has been highlighted  
in the context of LISA observations (see Vecchio 2004; Lang \& Hughes 2006) but no study has  
been devoted so far to ground-based observations.

In this \emph{Letter} we examine for the first time the potential for parameter estimation of  
spinning binary inspirals with ground-based interferometers, including twelve
physical parameters. Earlier studies (\emph{e.g.} Cutler \& Flanagan 1994, Poisson \& Will 1995, 
Van den Broeck \& Sengupta 2007) have estimated the theoretical accuracy with which some
of these parameters should be measured, without determining 
the parameters themselves. Also, (R\"{o}ver et al.\ 2006, 2007) have explored parameter 
estimation for non-spinning binaries. We focus on BH-NS binaries  
where spin effects are strongest 
(Apostolatos et al.\ 1994), while at the same time we are justified to ignore the NS spin.
We employ a newly developed Markov-chain Monte-Carlo (MCMC) algorithm (van der Sluys et al.\ 2008)
applied on spinning inspiral signals injected into synthetic ground-based noise and we derive posterior  
probability-density functions (PDFs) of all twelve signal parameters. 
We show that although sky position is degenerate 
when using two detectors, we can still determine the mass and 
spin parameters to reasonable accuracy.  With three detectors, the sky position 
and binary orientation can be fully resolved.
We show that our accuracies are good enough to associate an inspiral event with an electromagnetic
detection, such as a short gamma-ray burst (\emph{e.g.} Nakar 2007).

\section{Signal and observables}
\label{sec:GW}

In this \emph{Letter} we concentrate on the signal produced during the inspiral phase of two  
compact objects of masses $M_{1,2}$ in circular orbit. We  
focus on a fiducial BH-NS binary system with $M_1 = 10\,\Ms$ and $M_2 = 1.4\,\Ms$, so  
that we can ignore the NS spin. The BH spin $\mathbf{S}$
couples to the orbital angular momentum, leading to amplitude and phase modulation  
of the observed radiation due to the precession of the orbital plane during the observation. 
Here we model GWs by post-Newtonian (pN) waveforms at $^{1.5}$-pN order in phase and Newtonian 
amplitude.
We adopt the \emph{simple-precession} 
limit (Eqs.\,51, 52, 59 \& 63 in Apostolatos et al.\ 1994), appropriate  
for the single-spin system considered here. For simplicity (to speed up the  
waveform calculation), we ignore the \emph{Thomas-precession phase} (Apostolatos et al.\ 1994).
In this simple-precession approximation, the orbital angular momentum $\mathbf{L}$ and
spin $\mathbf{S}$ precess with the \emph{same} angular frequency 
around a fixed direction $\hat{\mathbf{J}}_0 \approx \hat{\mathbf{J}}$, where $\mathbf{J} = \mathbf{L} + \mathbf{S}$.
During the inspiral phase the spin misalignment
 $\theta_\mathrm{SL} \equiv \mathrm{arccos}({\bf \hat S} \cdot {\bf \hat L})$ and $S = |\mathbf{S}|$ 
are constant. These approximated waveforms retain (at the leading  
order) all the salient qualitative features introduced by the spins, while allowing us to  
compute the waveforms analytically, at great computational  
speed. While this approach is justified for exploration of GW astronomy and development of  
parameter-estimation algorithms, more accurate waveforms (\emph{e.g.} Kidder 1995;  
Faye et al.\ 2006; Blanchet et al.\ 2006) will be necessary for the analysis of real signals. 

A circular binary inspiral with one spinning compact object is described by a 12-dimensional parameter  
vector $\vec{\lambda}$. With respect to a fixed geocentric coordinate system our choice of  
independent parameters is:
\be
\vec{\lambda} = \{\Mc,\eta,\mathrm{R.A.},\mathrm{Dec},\cos\theta_{J_0},\phi_{J_0},\log{d_L},a_\mathrm{spin},\cos\theta_\mathrm{SL}, \phi_c,\alpha_c, t_c\},
\label{e:lambda}
\ee
where $\Mc = \frac{(M_1 M_2)^{3/5}}{(M_1 + M_2)^{1/5}}$ and $\eta = \frac{M_1 M_2}{(M_1 + M_2)^2}$ are  
the chirp mass and symmetric mass ratio, respectively; $\mathrm{R.A.}$ (right ascension)  
and $\mathrm{Dec}$ (declination) identify the source position in the sky;  
the angles $\theta_{J_0}$ and $\phi_{J_0}$ (defined in the range $\theta_{J_0} \in \left[-\frac{\pi}{2},\frac{\pi}{2} \right]$  
and $\phi_{J_0}\in \left[0, 2\pi \right[$) identify the unit vector $\hat{\mathbf{J}}_0$; $d_L$ is the  
luminosity distance to the source and $0 \le a_\mathrm{spin} \equiv S/M_1^2 \le 1$ is the  
dimensionless spin magnitude; $\phi_c$ and $\alpha_c$ are integration constants that specify  
the GW phase and the location of $\mathbf{S}$ on the precession cone, respectively, at 
the time of coalescence $t_c$. 

Given a network comprising $n_\mathrm{det}$ detectors, the data collected at the $a-$th  
instrument ($a = 1,\dots, n_\mathrm{det}$) is given by $x_a(t) = n_a(t) + h_a(t;\vec{\lambda})$,  
where $h_a(t;\vec{\lambda}) = F_{a,+}(t)\,h_{a,+}(t;\vec{\lambda}) + F_{a,\times}(t)\,h_{a,\times}(t;\vec{\lambda})$  
is the GW strain at the detector (see Eqs.\,2--5 in Apostolatos et al.\ 1994)
and $n_a(t)$ is the detector noise. The astrophysical signal  
is given by the linear combination of the two independent polarisations $h_{a,+}(t;\vec{\lambda})$  
and $h_{a,\times}(t;\vec{\lambda})$ weighted by the \emph{time-dependent} antenna beam  
patterns $F_{a,+}(t)$ and $F_{a,\times}(t)$. An example of $h_a$ for $\theta_\mathrm{SL}=20^\circ$  
and $a_\mathrm{spin}=0.1$ and $0.8$ is shown in panels a--b of Fig.\,\ref{fig:pdf}. In our  
analysis we model the noise in each detector as a zero-mean Gaussian,  stationary random  
process, with one-sided noise spectral density $S_a(f)$ at the initial-LIGO design sensitivity,  
where $f$ is the frequency. 

\section{Parameter estimation: Methods and results}
\label{sec:results}

The goal of our analysis is to determine the \emph{posterior} PDF of the unknown parameter  
vector $\vec{\lambda}$ in Eq.~(\ref{e:lambda}), given the data sets $x_a$ collected by a  
network of $n_\mathrm{det}$ detectors and the \emph{prior} $p(\vec{\lambda})$ on the  
parameters.  We use wide, flat priors (see Van der Sluys et al.\ 2008 for details).
Bayes' theorem provides a rigorous mathematical rule to assign such a probability: 
\be
p(\vec{\lambda}|x_a ) = \frac{p(\vec{\lambda}) \, {\cal L} (x_a|\vec{\lambda})}{p(x_a)}\,;
\label{e:jointPDF}
\ee
in the previous Equation
\be
{\cal L}(x_a|\vec{\lambda}) \propto 
\exp\left\{-2
\int_{f_l}^{f_h} \frac{\left| \tilde{x}_a(f) - \tilde{h}_a(f;\vec{\lambda})\right|^2}{S_a(f)}\,\mathrm{d}f
\right\}
\label{e:La}
\ee
is the \emph{likelihood function} of the data given the model, 
which measures the fit of the data to the model,
and $p(x_a)$ is the  
\emph{marginal likelihood} or \emph{evidence}; $\tilde x(f)$ stands for the Fourier  
component of $x(t)$. For multi-detector observations involving a network of detectors with  
uncorrelated noise --- this is the case of this paper, where we do not use the pair of  
co-located LIGO instruments 
--- we have $p(\vec{\lambda}|\{x_a; a  
= 1,\dots,n_\mathrm{det}\}) = \prod_{a=1}^{n_\mathrm{det}}\, p(\vec{\lambda}|x_a)\,.$

The numerical computation of the joint and \emph{marginalised} PDFs involves the  
evaluation of integrals over a large number of dimensions. Markov-chain Monte-Carlo  
(MCMC) methods (\emph{e.g.} Gilks et al.\ 1996; Gelman et al.\ 1997, and references therein)  
have proved to be particularly effective in tackling these numerical problems.
We have developed an adaptive (see Figueiredo \& Jain, 2002; Atchade \& Rosenthal 2003)  
MCMC algorithm, intended to explore the parameter  
space efficiently while requiring the least amount of tuning for the specific signal at  
hand; the code is an extension of the one developed by some of the authors to explore  
MCMC methods for non-spinning binaries (R\"{o}ver et al.\ 2006, 2007) and takes advantage  
of techniques explored by some of us in the context of LISA data analysis (Stroeer et al.\  
2007).  A summary of the methods used in our MCMC code has been published (Van der Sluys 
et al.\ 2008); more technical details will be provided elsewhere.

\begin{figure*} %For emulateapj
  %\figurenum{}
  \epsscale{1.0}
  \plotone{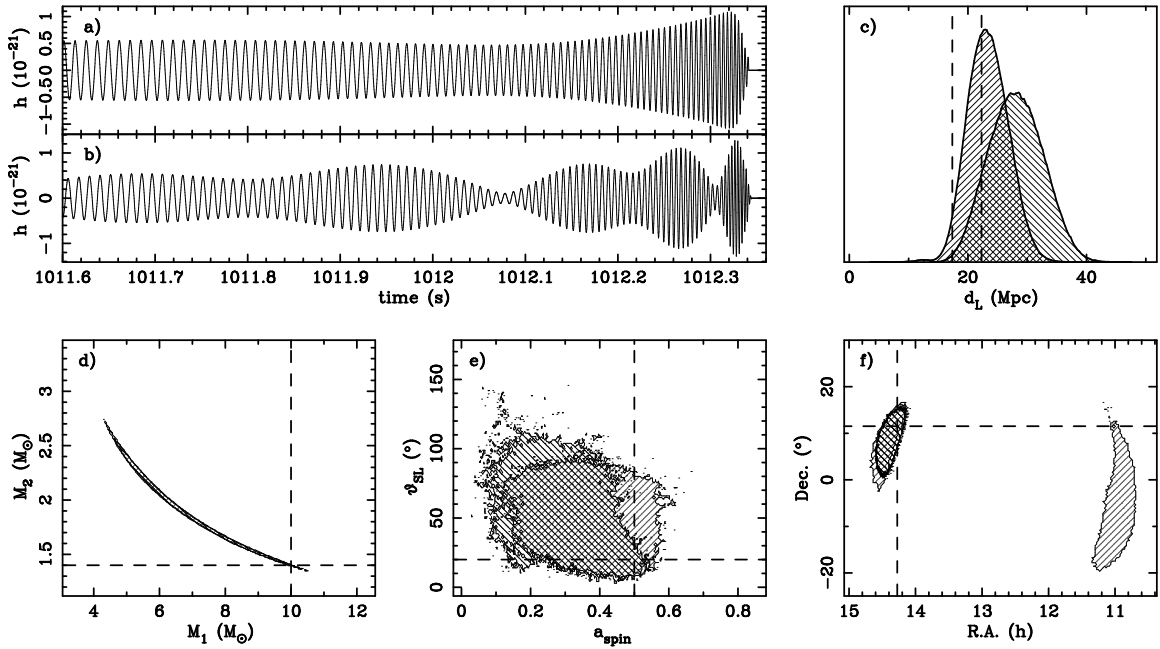} %For submission
  %\plotone{figures/waveform_pdfs.eps} %Local
  \figcaption{
    {\bf (a)} Part of the waveform from a source with $a_\mathrm{spin}=0.1$ and  
    $\theta_\mathrm{SL}=20^\circ$.   
    {\bf (b)} the same waveform, but for $a_\mathrm{spin}=0.8$. 
    {\bf (c)} Posterior PDF of the luminosity distance for a signal with $a_\mathrm{spin}=0.5$  
    and $\theta_\mathrm{SL}=20^\circ$, as determined with the signal of two (left PDF) and  
    three (right PDF) detectors. The dashed lines show the true distance, which is higher 
    for the three-detector case to obtain the same SNR.   
    {\bf (d--f)} Two-dimensional posterior PDF showing the 99\%-probability areas for 
    the same runs as (c), for the individual masses, 
    where the ellipses are aligned with the line of constant $\cal{M}$ 
    (d), the spin parameters (e) and the position in the sky (f). 
    The dashed lines display the true parameter values.
    Upward and downward hashes show the result for two and three detectors respectively 
    in panels {\bf (c--f)}. 
    \label{fig:pdf}
  }
%\end{figure}  %For aastex
\end{figure*}

Here we present results obtained by adding a signal in simulated initial-LIGO noise 
and computing the posterior PDFs with MCMC techniques for a fiducial  
source consisting of a $10\,M_\odot$ spinning BH and a $1.4\,M_\odot$ non-spinning NS  
in a binary system with a signal-to-noise ratio (SNR) of 17.0 for the network of 2 or 3 detectors
(obtained by scaling the distance). We consider a number of cases for which we  
change the BH spin magnitude ($a_\mathrm{spin} = 0.0, 0.1, 0.5, 0.8$) and the angle between  
the spin and the orbital angular momentum ($\theta_\mathrm{SL} = 20^\circ, 55^\circ$);   
the remaining ten parameters, including source position and orientation of the total  
angular momentum, are kept constant ($R.A. = 14.3h$, Dec. $= 12^\circ$, $\theta_{J_0} = 4^\circ$
and $\phi_{J_0} = 289^\circ$ for this study). For each of the seven ($a_\mathrm{spin}$, $\theta_\mathrm{SL}$)  
combinations (six for finite spin, one for zero spin), we run the analysis using the 
data from (i) the 4-km LIGO detector at Hanford (H1) and the Virgo detector near Pisa
($n_\mathrm{det}=2$), and (ii) the two LIGO 4-km interferometers (H1 and L1) and the Virgo 
detector ($n_\mathrm{det}=3$). This results in a total of 14 signal cases explored in this study. 
The MCMC analysis that we carry out on each data set consists of 5 separate serial chains, 
each with a length of $3.5\times10^6$ iterations ($n_\mathrm{det}=2$) or $2.5\times10^6$ 
iterations ($n_\mathrm{det}=3$),
sampled after a \emph{burn-in} period (see \emph{e.g.} Gilks et al.\ 1996) that is determined
automatically as follows: we determine the absolute maximum likelihood $L_\mathrm{max}$ 
that is obtained in any of the five chains, and for each chain include all the iterations 
{\it after} the chain has reached a likelihood value of $L_\mathrm{max}-2$.
Each chain starts at offset ({\it i.e.}, non-true) parameter values. The starting values
for chirp mass and the time of coalescence are drawn from a Gaussian distribution 
centred on the true parameter value, with a standard deviation of about $0.1\,M_\odot$ and 
30\,ms respectively.  The other ten parameters are drawn randomly from the allowed ranges.
Multiple chains starting from offset parameters and locking on to the same values for the
parameters and likelihood provide convincing evidence of convergence in a blind analysis.
Our MCMC code needs to run for typically one week to show the first results and 10--14 days
to accumulate a sufficient number of iterations for good statistics, each serial chain using
a single 2.8\,GHz CPU.
An example of the PDFs obtained for a signal characterised by $a_\mathrm{spin}=0.1$ and  
$\theta_\mathrm{SL}=20^\circ$ is shown in panels c--f of Fig.\,\ref{fig:pdf}, for the cases  
of 2 and 3 detectors; the PDFs for $M_1$ and $M_2$ in Fig.\,\ref{fig:pdf}d are constructed  
from those obtained for $\Mc$ and $\eta$.

%\begin{figure}  %For aastex
 %For emulateapj

In order to evaluate the parameter-estimation accuracy we compute probability intervals; in  
Table\,\ref{tab:deltas} we report the 90\%-probability interval for each of the parameters,  
defined as the smallest range for which the posterior probability of a given parameter to be  
in that range is 0.9.  For the two-dimensional cases (position and orientation)
we quote the smallest \emph{area} that contains 90\% of the probability. 
From the 140 marginalised PDFs considered here (ignoring the 
derived parameters $M_1$, $M_2$ and combining R.A., Dec as position and $\theta_\mathrm{J0}$, 
$\phi_\mathrm{J0}$ as orientation), the true parameter values lie outside the $90\%$ probability 
range in 27 cases, marked with superscripts in Table\,\ref{tab:deltas}. For  
the 21 cases marked with $^a$ the true parameter is within the 99\%-probability range,
the 6 cases marked with $^b$ lie outside the 99\% but inside the 100\% range.

%\begin{deluxetable}{lllllllllllllllllll} %For aastex
\begin{deluxetable*}{lllllllllllllllllll} %For emulateapj
  %\tabletypesize{\scriptsize} %For aastex
  %\tablewidth{0pt}
  \tablecolumns{19}
  \tablecaption{
    Injection details and widths of the 90\%-probability intervals of the MCMC runs described in the text
    \label{tab:deltas}
  }
  \tablehead{
    \colhead{$n_\mathrm{det}$$\!\!\!$} & \colhead{$a_\mathrm{spin}$$\!\!\!$} & \colhead{$\theta_\mathrm{SL}$$\!\!\!$} & \colhead{$d_\mathrm{L}$$\!\!\!$} & 
    \colhead{~} &
    \colhead{$M_1$$\!\!\!$} & \colhead{$M_2$$\!\!\!$} &
    \colhead{~} &
    \colhead{$\cal{M}$$\!\!\!$} & \colhead{$\eta$$\!\!\!$} & \colhead{$t_\mathrm{c}$$\!\!\!$} & \colhead{$d_\mathrm{L}$$\!\!\!$} & 
    \colhead{$a_\mathrm{spin}$$\!\!\!$} & \colhead{$\theta_\mathrm{SL}$$\!\!\!$} & 
    \colhead{$\phi_\mathrm{c}$$\!\!\!$} & 
    \colhead{$\alpha_\mathrm{c}$$\!\!\!$} & 
    \colhead{~} &
    \colhead{Pos.$\!\!\!$} & \colhead{Ori.$\!\!\!$} \\
    & & ($^\circ$) & (Mpc) & 
    &
    (\%) & (\%) & 
    &
    (\%) & (\%) & (ms) & (\%) & 
    & ($^\circ$) & 
    ($^\circ$) &  ($^\circ$)  &  
    &
    ($^{\circ^2}$) &  ($^{\circ^2}$) 
  }
  \startdata
  2  &     0.0  &       0  &    16.0  &  &    95       &    83       &  &   2.6   &   138   &    18   &    86   &  0.63   &---   &   323   &---   & &   537   & 19095  \\
  2  &     0.1  &      20  &    16.4  &  &   102       &    85       &  &   1.2   &    90   &    10   &    91   &  0.91   &   169   &   324   &   326$^a$   & &   406   & 16653  \\
  2  &     0.1  &      55  &    16.7  &  &    51       &    38       &  &  0.88   &    59   &   7.9   &    58   &  0.32   &   115   &   322   &   326   & &   212   &  3749  \\
  2  &     0.5  &      20  &    17.4  &  &    53$^b$   &    42$^a$   &  &  0.90   &    50$^b$   &   5.4   &    46$^a$   &  0.26   &    56   &   330   &   301$^b$   & &   111$^a$   &  3467$^a$  \\
  2  &     0.5  &      55  &    17.3  &  &    31       &    24       &  &  0.62   &    41   &   4.9   &    21   &  0.12   &    24   &   323   &   269$^a$   & & 19.8   &   178$^a$  \\
  2  &     0.8  &      20  &    17.9  &  &    54$^a$   &    42$^a$   &  &  0.86$^a$   &    54$^a$   &   6.0   &    56   &  0.16   &    25$^a$   &   325   &   319   & &  104$^a$   &  1540  \\
  2  &     0.8  &      55  &    17.9  &  &    21       &    16       &  &  0.66   &    29   &   4.7   &    22   &  0.15   &    15   &   320   &   323   & & 22.8   &   182$^a$  \\
  & & & &  &  & &  &   & & & & & & & &  & & \\
  3  &     0.0  &       0  &    20.5  &  &   114       &    90       &  &   2.6   &   119   &    15   &    69   &  0.98$^b$   &---   &   325   &---   & &  116   &  4827  \\
  3  &     0.1  &      20  &    21.1  &  &    70       &    57       &  &  0.92   &    72   &   7.0   &    60   &  0.49   &   160   &   321   &   322$^a$   & &  64.7   &  3917  \\
  3  &     0.1  &      55  &    21.4  &  &    62       &    48       &  &  0.93   &    68   &   6.2   &    51   &  0.52   &   123   &   325   &   308$^a$   & & 48.7   &   976  \\
  3  &     0.5  &      20  &    22.3  &  &    54$^b$   &    44$^a$   &  &  0.89$^a$   &    48$^b$   &   3.3   &    52   &  0.28$^a$   &    69   &   318   &   229$^b$ &  &  28.8   &   849  \\
  3  &     0.5  &      55  &    22.0  &  &    33       &    25       &  &  0.62   &    43   &   4.6   &    23$^a$   &  0.14   &    27   &   322   &   324   & &  20.7   &   234$^a$  \\
  3  &     0.8  &      20  &    23.0  &  &    53$^b$   &    41$^a$   &  &  0.85$^a$   &    52$^b$   &   3.8   &    55   &  0.17   &    23$^a$   &   320   &   327$^a$  & &  36.4$^a$   &   645  \\
  3  &     0.8  &      55  &    22.4  &  &    30       &    22       &  &  0.86   &    40   &   5.0   &    26   &  0.21   &    21   &   322   &   323   & &  27.2   &   288 
  \enddata
  \tablenotetext{~}{$^a$ the true value lies outside the 90\%-probability range; $^b$ idem, outside the 99\%-probability range, but inside the 100\% range}
%\end{deluxetable} %For aastex
\end{deluxetable*} %For emulateapj

We find that most of these outliers are caused by a degeneracy between the mass and spin 
parameters.  A parameter set with different values for $\cal{M}$, $\eta$, $a_\mathrm{spin}$ 
and $\theta_\mathrm{SL}$ can produce a waveform that is almost identical to the signal we
injected.  For the chirp mass and spin parameters, the distance between the two degenerate
regions is relatively small. 
However, for the mass ratio $\eta$, these two regions ($\eta\approx0.11$, the injected value
and $\eta\approx0.2$) are far apart and seem disconnected. 
A comparison of waveforms from the two degenerate regions demonstrates that their overlap is so
high (match $>99.5\%$) that it would be impossible to tell which is the true signal even at 
high SNR.  This degeneracy could be physical or could be caused by the simplified waveform model;
further investigation is warranted.

For a detection with two interferometers, the sky position and binary orientation are 
degenerate; for low spin, our PDFs show an incomplete ring in the sky where the source might 
be.  When the BH spin increases, the allowed sky location shrinks appreciably until mere arcs 
are left (Fig.\,\ref{fig:pdf}f).
For intermediate and high spin, and $\theta_\mathrm{SL}=55^\circ$, we typically find only one
such arc, reducing the sky position to several degrees (Table\,\ref{tab:deltas}).

Even with two detectors the source parameters can be measured at 
astrophysically interesting levels when sufficient spin is present, including distance, individual  
masses, spin magnitude and tilt angle; 
for $a_\mathrm{spin}=0.5$ or more, the typical uncertainty in the sky position is
a few tens of square degrees, the distance is determined with 20--60\% accuracy and the timing 
accuracy is 6\,ms or better. 

The accuracy of the parameter determination is affected by the number of 
detectors used, a result well established in studies of inspirals of non-spinning 
compact objects (\emph{e.g.} Jaranowski, P., \& Krolak 1994; Pai et al.\ 2001; Cavalier et al.\ 2006; 
R\"{o}ver et al.\ 2007).
Unlike some other studies, we keep the SNR of the detector network constant; when a third detector is added, the
distance to the source is increased (Fig.\,\ref{fig:pdf}c).  This way, we see the effect of the additional information
that is provided by the extra interferometer and eliminate the effect of the higher SNR.
Table\,\ref{tab:deltas} shows that the effect on the uncertainty in the mass and spin 
parameters is marginal when adding a third interferometer to the network.
The uncertainty in the distance and time of coalescence 
decreases typically by 20--25\% when using three detectors, but the largest effect is on the
accuracy for sky position and binary orientation; Table\,\ref{tab:deltas}
shows that the (two-dimensional) uncertainties in these parameters decrease by 50\% and
40\% respectively on average.
%See ~/work/GW/programs/MCMC/output/papers/ApJL_01/draft_0806/table

The parameter-estimation accuracy also depends strongly on the actual spin 
parameters of the system: as a general trend, the larger $a_\mathrm{spin}$ and $\theta_\mathrm{SL}$,  
the stronger the modulations in the waveform induced by precession, and the more information
is coded up in the waveform.  
When we divide our simulations into low spin ($a_\mathrm{spin}=0.0,0.1$) and high spin 
($a_\mathrm{spin}=0.5,0.8$) cases, we find that the uncertainties in the high-spin case are smaller
by 40--60\% for the masses, time of coalescence and distance, by 65--70\% for the 
spin parameters and by 80--90\% for the sky position and binary orientation.
However, the width of the 90\%-probability interval is in fact not strictly monotonic as a function 
of $a_\mathrm{spin}$ and $\theta_\mathrm{SL}$ (Table\,\ref{tab:deltas}).
The increasingly complex structure of the likelihood 
function and stronger correlations amongst different parameters for higher spin have an important 
effect on the sampling efficiency of the MCMC.

Earlier studies
(\emph{e.g.} Cutler \& Flanagan (1994, their Tables\,II \& III and Fig.\,7);
Poisson \& Will (1995, their Table\,II); Van den Broeck \& Sengupta (2007, their Table\,III))
have reported on the theoretical accuracy of parameter estimation. 
These explorations are based on the Fisher 
matrix, which yields the expected uncertainty (for unimodal 
distributions), without actually estimating the parameter values themselves.  The quoted accuracies 
for masses and the time and phase of coalescence are typically better than or similar to the 
values in our Table\,\ref{tab:deltas}.  We have been able to estimate distance, sky position and 
binary orientation to better accuracy than suggested in these studies.

\section{Conclusions}
\label{sec:concl}

We have explored for the first time the 
parameter estimation
of all physical parameters --- including masses, spin, 
distance, sky location and binary orientation --- on ground-based gravitational-wave
observations of binary inspirals with spinning compact objects. 
We show that for two detectors and sufficient spin ($a_\mathrm{spin}\geq0.5$) or
for three detectors, the obtained accuracy in sky position, distance and time of 
coalescence is good enough to allow the identification of electromagnetic 
counterparts of compact-binary mergers, \emph{e.g.} short gamma-ray bursts (Nakar 2007).
A direct measurement of mass, spin, distance and orientation can be obtained from inspiral GWs, which
is notoriously difficult for electromagnetic observations. 

The analysis presented here is the first step of a more detailed  
study that we are currently carrying out, exploring a much larger parameter space,  
developing techniques to reduce the computational cost of these simulations, and  
testing the methods with actual LIGO data. The waveform model used here, though adequate for  
exploratory studies, is not sufficiently accurate for the analysis of real detections, 
and we are finalising the implementation of a more realistic waveform. Simulations
with this improved waveform may also shed light on the degeneracy between mass
and spin parameters discussed in Sect.\,\ref{sec:results}, and may improve the 
accuracy of our parameter estimation appreciably (\emph{e.g.} Van den Broeck \& 
Sengupta 2007).
Finally, we intend to further develop our Bayesian approach into one of the standard  
tools that can be included in the analysis pipeline used for the processing of the  
`science data' collected by ground-based laser interferometers.

\acknowledgments 

This work is partially supported by a Packard Foundation Fellowship, a NASA BEFS grant  
(NNG06GH87G), and a NSF Gravitational Physics grant (PHY-0353111) to VK; NSF Gravitational  
Physics grant PHY-0553422 to NC; Royal Society of New Zealand Marsden Fund grant UOA-204  
to RM and CR; UK Science and Technology Facilities Council grant to AV; Computations were  
performed on the Fugu computer cluster funded by NSF MRI grant PHY-0619274 to VK.

\end{document}